\documentclass[aps,floatfix,pra,reprint,superscriptaddress]{revtex4-2}

\usepackage{amsmath}
\usepackage{amssymb}
\usepackage{amsthm}
\usepackage{bbm}
\usepackage{color}
\usepackage{float}
\usepackage[T1]{fontenc}
\usepackage{graphicx}
\usepackage{hyperref}
\usepackage[utf8]{inputenc}
\usepackage{orcidlink}
\usepackage{physics}
\usepackage{times}
\usepackage{txfonts}
\usepackage{xcolor}
\usepackage{color,soul}
\usepackage{lineno}
\hypersetup{colorlinks=true,linkcolor=blue,citecolor=blue,filecolor=blue,urlcolor=blue,breaklinks=true}

\theoremstyle{definition}

\def\be{\begin{equation}}
\def\ee{\end{equation}}

\def\bc{\begin{center}}
\def\ec{\end{center}}
\def\bal{\begin{align}}
\def\eal{\end{align}}

\def\etal{\textit{et al. }}

\newcommand{\defis}{
  Departamento de Física,
  Universidade Estadual de Ponta Grossa
  84030-900 Ponta Grossa, Paraná, Brazil
}

\newcommand{\demat}{
  Departamento de Matemática e Estatística,
  Universidade Estadual de Ponta Grossa,
  84030-900 Ponta Grossa, Paraná, Brazil
}
\newcommand{\qpqi}{
  QPQI Group,
  Universidade Estadual de Ponta Grossa,
  84030-900 Ponta Grossa, Paraná, Brazil
}

\newcommand{\orcidthiago}{\orcidlink{0009-0001-1654-0330}}
\newcommand{\orcidalison}{\orcidlink{0000-0003-3552-8780}}
\newcommand{\orcidantonio}{\orcidlink{0000-0002-1521-9342}}
\newcommand{\orcidfabiano}{\orcidlink{0000-0001-5383-6168}}

\begin{document}

\title{
Optical perspective on the time-dependent Dirac oscillator
}

\author{Thiago T. Tsutsui\orcidthiago}
\email{takajitsutsui@gmail.com}
\affiliation{\qpqi}

\author{Alison A. Silva\orcidalison}
\email{alisonantunessilva@gmail.com}
\affiliation{\qpqi}

\author{Antonio S. M. de Castro\orcidantonio}
\email{asmcastro@uepg.br}
\affiliation{\qpqi}
\affiliation{\defis}

\author{Fabiano M. Andrade\orcidfabiano}
\email{fmandrade@uepg.br}
\affiliation{\qpqi}
\affiliation{\demat}

\date{\today}

\begin{abstract}
The Dirac oscillator is a relativistic quantum system, characterized by its linearity in both position and momentum. Moreover, considering $(1{+}1)$ and $(2{+}1)$ dimensions, the system can be mapped onto the Jaynes-Cummings and anti-Jaynes-Cummings models, as illustrated in an exact manner by Bermudez \etal [\href{ https://doi.org/10.1103/PhysRevA.76.041801}{Phys. Rev. A 76, 041801(R) (2007)}]. Using the optical counterparts of the Dirac oscillator, we analyze an extension of the model that incorporates a time-dependent frequency. We focus on the consequences of these time modulations on the angular momentum observables and spin-orbit entanglement. Noticeable changes in the \emph{Zitterbewegung} are found. We show that a specific choice of time dependence yields aperiodic evolution of the observables, whereas an alternative choice allows analytical solutions.\\

\noindent
\textbf{Front. Phys. 14, 1772868 (2026);}
doi: \href{http://dx.doi.org/10.3389/fphy.2026.1772868}
{10.3389/fphy.2026.1772868}
\end{abstract}

\maketitle

\section{Introduction}
\label{sec:introduction}

In relativistic quantum mechanics, the Dirac equation
\cite{dirac1928a,dirac1928b} stands as the framework for describing
fermions, linearly expressing the relationship between energy and linear
momentum.
Given that the quantum harmonic oscillator (HO), derived from the
Schr\"odinger equation, is a cornerstone in physics, it is natural to
consider a quantum HO that emerges from the Dirac equation.
Discerned by its linearity in position and momentum, the Dirac
oscillator (DO) fulfills this role, as it reduces to a quantum HO with
strong spin-orbit coupling in the nonrelativistic limit.
The model was originally proposed by Itô \emph{et al}. \cite{ITO1967}
and subsequently explored by other researchers \cite{Cook1971,Ui1984}.
However, only with Moshinsky and Szczepaniak \cite{MOCHINSKY1989} the
name ``Dirac oscillator'' was coined, along with the introduction of the
non-minimal coupling associated with the model.
As discussed in Ref. \cite{Martinez-y-Romero1995}, one can regard the
physical picture of DO as a neutral particle interacting with a static
linear electric field.
From a mathematical perspective, the model exhibits a Lie algebra
\cite{Quesne1990} and supersymmetry properties \cite{Bentez1990}.

Furthermore, the DO has been investigated from various perspectives,
including many-body physics \cite{Moshinsky1991},
the Dunkl formalism \cite{Ojeda-Guillen2020,Hamil2022,Hamil2022a}, 
the Foldy-Wouthuysen
transformation \cite{Moreno1989}, thermal properties \cite{Boumali2013},
quantum deformations \cite{Andrade2014a,Andrade2014b} and fractional
calculus \cite{Korichi2022}.
In terms of applications, the Dirac oscillator can be employed to
analyze the electronic dynamics of two-dimensional materials, such as
graphene \cite{Sadurni2010}, and to model a quark confinement potential
in quantum chromodynamics \cite{Moreno1989}.
Experimentally, the DO was implemented in a microwave setup using a
tight-binding correspondence \cite{Franco-Villafane2013}, while
simulations and experimental realizations have also been pursued in
other platforms \cite{BERMUDEZ2007,Longhi2010,Zhang2018}.
For an overview of the DO, we suggest Ref. \cite{Sadurni2011}.

Relationships between the DO and optical systems, which are more robust
regarding practical verification, have been identified
\cite{Rozmej1999}.
Bermudez \etal \cite{BERMUDEZ2007} demonstrated an exact mapping
between the $(2{+}1)$ DO  and the paradigmatic Jaynes-Cummings (JC)
model \cite{Jaynes1963}, which depicts the interaction of a two-level
atom with one mode of a quantized cavity.
Applying the mapping, the authors proposed an experiment for the
experimental study of the expectation value of the angular momentum
observables.
Since then, other works have further advanced this connection \cite{Bermudez2008a,Obada2019,UHDRE2022,Dagoudo2024}.

To contribute to this topic, we investigate a generalization of the DO
with a time-dependent frequency, inspired by the time-dependent JC
(TDJC) model \cite{Prants1992,Schlicher1989,Bartzis1992,Wilkens1992,
Joshi1993,Prants1997,Fang1998,Dasgupta1999,Larson2021},
where the parameters, otherwise constant, may present time dependence.
The DO with a time-dependent frequency was studied in
Ref. \cite{Sobhani2015} within the Lewis-Riesenfeld framework
\cite{Lewis1969,Landim2000,DeCastro2003,Zhang2007,
Benchikha2024,Benchikha2025,Lutfuoglu2025}.
The path integral method \cite{Poulter1994,Gosselin1998,Benchikha2025a}
and the Floquet formalism \cite{Rodriguez-Lopez2014,Junk2020,Wu2025}
provide other approaches to time-dependent quantum mechanical problems.
Here, however, we approach the problem from a different perspective --
namely, a quantum optics approach.
We focus our analysis and notation on the $(2{+}1)$ DO; however, as we
will demonstrate, our results extend to the $(1{+}1)$ DO, and to the
anti-Jaynes-Cummings (AJC) mapping as well.
To account for the impact of this generalization, we examine the
expectation values of angular momentum observables and spin-orbit
entanglement, measured according to the von Neumann entropy
\cite{VonNeumann1927b}.
A similar ``optical perspective'' can be seen in Ref. \cite{Obada2019},
with the DO coupled to an external isospin field.

We insert the frequency as the parameter to depend on time to draw
comparisons with specific works in the TDJC literature
\cite{Schlicher1989,Prants1992,Joshi1993,Fang1998,Dasgupta1999,DeCastro2023}.
For two specific physical scenarios in the optical analog of the DO, we
employ two types of modulation: a trigonometric
\cite{Schlicher1989,Fang1998} and an exponential \cite{Prants1992} one.
Both cause noticeable changes in the \emph{Zitterbewegung}
\cite{SCHRöDINGER1930}, significantly altering the behavior of the
angular momentum observables and entanglement.
The trigonometric modulation may induce aperiodic behavior, whereas the
exponential modulation allows for analytical solutions.
We conduct simulations for both an initial number state and a coherent
state of the orbital angular momentum. 
To our knowledge, such an analysis has not been performed for the DO,
rendering our results novel in the context of relativistic quantum
mechanics.

The time-dependent extension of the model may allow for a broader range
of physical situations in which the DO can be emulated and potentially
verified.
For example, the exponential modulation can model transient effects in a
simulation.
Furthermore, in cavity quantum electrodynamics \cite{Haroche2006}, the
semiclassical atomic motion in a standing-wave mode is translated as a
trigonometric light-matter coupling.

This paper is organized as follows.
In Sec. \ref{sec:dirac_osc}, we introduce the model and its mapping to
optical systems.
In Sec. \ref{sec:time_dep}, we present the results for the
time-dependent extension of the DO, including the analytical solution
for the exponential modulation, the behavior of the expectation values
of the angular momentum observables, and the dynamics of spin-orbit
entanglement.
Our conclusions are summarized in Sec. \ref{sec:conc}.
In the appendix, we show how our results can be extended to the AJC
mapping.

\section{Dirac Oscillator}
\label{sec:dirac_osc}

In relativistic quantum mechanics \cite{Greiner2000}, the Dirac equation
\cite{dirac1928a,dirac1928b} provides a theoretical formalism for
describing spin-$1/2$ particles, known as fermions.
The DO is obtained by the introduction of the non-minimal coupling
$\mathbf{p}\to\mathbf{p}\pm im\omega\beta\mathbf{r}$ in the Dirac
equation  \cite{MOCHINSKY1989,Sadurni2011,UHDRE2022}, i.e.
\begin{equation}
  \label{eq:do_3+1}
  i\hbar\frac{\partial}{\partial t}\ket{\Psi(t)}
  =\left[c\boldsymbol{\alpha}\cdot(\mathbf{p}
    \pm im\omega\beta\mathbf{r})
    +\beta mc^{2}\right]\ket{\Psi(t)}.
\end{equation}
The state $\ket{\Psi(t)}$ represents a four-component spinor, $c$
denotes the speed of light, $m$ the rest mass, $\mathbf{p}$ the linear
momentum, $\omega$ the oscillator frequency, and $\mathbf{r}$ the
position.
The matrices $\boldsymbol{\alpha}$ and $\beta$, known as the Dirac
matrices, obey the Clifford Algebra \cite{Garling2011}, characterized by
the anti-commutation relations
\begin{align}
  \{\alpha_{j},\alpha_{k}\} = {} & 2\delta_{jk},\\
  \{\alpha_{j},\beta\} = {} & 0.
\end{align}
Reflecting an important aspect of the standard Dirac equation
(retroactively obtained by setting $\omega = 0$ in
Eq. \eqref{eq:do_3+1}),  the DO exhibits a linear relationship between
position and momentum.
Additionally, in the nonrelativistic limit, we recover the Schr\"odinger
equation for the HO,  with a strong spin-orbit coupling term
\cite{MOCHINSKY1989}. 

In the following considerations, we adopt only the positive sign in
Eq. \eqref{eq:do_3+1} for simplicity.
The specific choice of the positive sign will be clarified further
below.
In the following, we rewrite the DO Hamiltonian in $(1{+}1)$ and
$(2{+}1)$ dimensions, establishing its correspondence with the JC model.
In the appendix, we discuss the alternative choice of sign and the
resulting differences.

\subsection{$(1{+}1)$ Dirac oscillator}
\label{sec:1_do}

First, we consider the $(1{+}1)$ dimensions scenario.
This case was experimentally realized in a tight-binding system
\cite{Franco-Villafane2013}, using the positive sign in
Eq. \eqref{eq:do_3+1}.

We consider that the movement limits itself to the $x$-axis.
Under this constraint, $|\Psi(t)\rangle$ can be described by a
two-component spinor and  the Dirac matrices reduce to the Pauli
matrices, $\alpha_x = \sigma_{x}$ and $\beta = \sigma_{z}$,
and the Hamiltonian of Eq. \eqref{eq:do_3+1} simplifies to \cite{Sadurni2011,Dagoudo2024}
\begin{equation}
  \label{eq:do1}
  H_{1}=c\sigma_{x}\left(p_{x}
    + i m\sigma_{z}\omega x\right)+\sigma_{z}mc^{2}.
\end{equation}
Following the nonrelativistic approach to the  HO \cite{SAKURAI2020}, we
can write the position and momentum as functions of the creation and
annihilation operators
\begin{align}
  \label{eq:r_i_pi_a_a_dagg}
  r_i = {} & \sqrt{\frac{\hbar}{2m\omega}}(a_{r_i}^{\dagger}+a_{r_i}), \\
  p_i = {} & i\sqrt{\frac{m\omega \hbar}{2}}(a_{r_i}^{\dagger}-a_{r_i}),
\end{align}
where, in the $(1{+}1)$ case, $r_i=x$.
Additionally, we employ the Pauli matrix form of $\sigma_i$,
\begin{equation} \label{eq:pauli_matrices}
\sigma_{x}=\begin{pmatrix}0 & 1\\
1 & 0
\end{pmatrix},\quad\sigma_{y}=\begin{pmatrix}0 & -i\\
i & 0
\end{pmatrix},\quad\sigma_{z}=\begin{pmatrix}1 & 0\\
0 & -1
\end{pmatrix}.
\end{equation}
Thus, we can rewrite $H_1$ as
\begin{equation}
    H_{1}=\begin{pmatrix}mc^{2} & -imc^2\sqrt{2\xi}a_{x}\\
imc^2\sqrt{2\xi}a_{x}^{\dagger} & -mc^{2}
\end{pmatrix},
\end{equation}
where we have introduced the relativistic parameter
\begin{equation} \label{eq:relativisitic_par}
    \xi=\frac{\hbar \omega}{mc^2}.
\end{equation}

Employing the number operator basis
\cite{BERMUDEZ2007,UHDRE2022,Dagoudo2024},
\begin{equation}
    \left|n_{x}\right\rangle =\frac{1}{\sqrt{n_{x}!}}(a_{x}^{\dagger})^{n_{x}}\left|0\right\rangle,
\end{equation}
with $n_x=0,1,2\dots$ representing the eigenvalues of the number
operator $N_x=a_x^\dagger a_x$,
we can find the energy eigenvalues for the model
\begin{equation} \label{eq:energy_1+1}
    E_{1}^{\pm}=\pm mc^{2}\sqrt{2\xi(1+n_{x})+1}.
\end{equation}
The double sign accounts for the positive and negative energy spectra
arising from the Dirac equation.
Furthermore, with the ladder operators
\begin{equation} \label{eq:ladder}
    \sigma_+=\begin{pmatrix}0 & 1\\
0 & 0
\end{pmatrix},
\qquad\sigma_-=\begin{pmatrix}0 & 0\\
1 & 0
\end{pmatrix},
\end{equation}
we can write the $(1{+}1)$ DO Hamiltonian as
\begin{equation} \label{eq:do_1_map}
  H_1=  \hbar (g_1 a_{x}^{\dagger}\sigma_{-}
  + g_1^* a_{x}\sigma_{+})+mc^{2}\sigma_{z},
\end{equation}
where the $(1{+}1)$ coupling is defined as
\begin{equation}
  g_1
  = \frac{ imc^2\sqrt{2\xi}}{\hbar}
  =\frac{ic}{\hbar}\sqrt{2 m\omega\hbar}.
\end{equation}
Equation \eqref{eq:do_1_map} represents a JC-like Hamiltonian
\cite{Sadurni2011,Dagoudo2024}.
In this context, the optical counterparts of $g_1$ and $mc^2$ correspond
to the atom–field coupling strength and detuning, respectively.
Choosing the negative sign earlier would have yielded an AJC-like
interaction, featuring terms such as $a_{x}\sigma_{-}$ and
$a_{x}^\dagger\sigma_{+}$.
Even if less intuitive, there is nothing intrinsically wrong with them;
however, the JC interactions appear to be more experimentally feasible
for simulation purposes.
Villafañe \etal \cite{Franco-Villafane2013} defined the condition
$mc^2\to0$ while $m\hbar\omega\to \mbox{const.}\neq0$ as the
``Weyl oscillator'' limit, associating it with the massless limit of the
Dirac equation \cite{weyl1929}.
This is equivalent to an on-resonance JC system.

The DO-JC mapping can be understood from the fact that the DO
interaction couples fermionic ($\sigma_\pm$) and bosonic ($a_x$ and
$a_x^\dagger$) degrees of freedom.
We may generically refer to the fermionic degree of freedom as a
two-level system (TLS), parameterized by the Pauli matrices.
Strictly speaking, spin is absent in the Dirac oscillator in $(1{+}1)$
dimensions, but the underlying algebra remains the same.
The actual spin operator in the $z$-direction emerges explicitly in the
$(2{+}1)$ case \cite{Sadurni2011}.
Furthermore, the bosonic degree of freedom corresponds to a linear HO
\cite{Cao2020}.
In the context of JC physics, this reflects the interplay between a
two-level atom and a single-mode cavity field.
It is worth noting that $g_1$ is proportional to the relativistic
parameter $\xi$; thus, in the nonrelativistic limit, the spin and the
HO degrees of freedom decouple.
The DO-JC connection was first pointed out by Rozmej and Arvieu in a
series of works \cite{Arvieu1994,Arvieu1995,Rozmej1996,Rozmej1999}.

In the JC model, there exists an operator, the excitation number, which
commutes with the Hamiltonian and, consequently, is a conserved quantity
\cite{Klimov2009}.
The existence of this operator restricts the dynamics to a $2{\times}2$
subspace and allows analytical solutions in the standard scenario
\cite{Klimov2009}.
In the $(1{+}1)$ DO notation \cite{Sadurni2011}, we define it as
\begin{equation}
  \label{eq:conserved}
    I_1=N_x+\frac {1}{2}\sigma_z,
\end{equation}
responsible for a $U(1)$ symmetry in the model
\cite{Sadurni2011,Larson2021}.

\subsection{$(2{+}1)$ Dirac oscillator}
\label{sec:2_do}

Now, we restrict the movement to the $xy$ plane, considering  $(2{+}1)$
dimensions \cite{Villalba1994}, with $\ket{\Psi(t)}$ again describing a
two-component spinor.
In this scenario, the DO Hamiltonian becomes
\begin{align}
  \label{eq:do_2}
  H_2  = {}
  &
     c\sigma_{x}\left(p_x + i m\sigma_{z}\omega x \right)
    +c\sigma_{y}\left(p_{y}+i m\sigma_{z}\omega
    y\right)+\sigma_{z}mc^{2} \nonumber
   \\
  =  {} 
  &
     H_1 +c\sigma_{y}\left(p_{y}-i m\sigma_{z}\omega y\right),
\end{align}
reducing to Eq. \eqref{eq:do1} when $p_y=y=0$.

We shall proceed with the same procedure as in the preceding subsection,
but with the introduction of the chiral operators
\begin{align}
  \label{eq:chiral_operators}
  a_{r}
  = {} & \frac{1}{\sqrt{2}}\left(a_{x}-i a_{y}\right),
  & \quad
    a_{r}^{\dagger}= {} & \frac{1}{\sqrt{2}}\left(a_{x}^{\dagger}+
                          ia_{y}^{\dagger}\right), \\
  a_{l}
  = {} & \frac{1}{\sqrt{2}}\left(a_{x}+i a_{y}\right),
  & \quad
    a_{l}^{\dagger}= {} &\frac{1}{\sqrt{2}}\left(a_{x}^{\dagger}-i
                          a_{y}^{\dagger}\right).
\end{align}
Employing Eqs. \eqref{eq:r_i_pi_a_a_dagg}, \eqref{eq:pauli_matrices} and
\eqref{eq:chiral_operators}, the Hamiltonian in matrix form is written as
\begin{equation}
  H_{2}=\begin{pmatrix}mc^{2} & -imc^22\sqrt{\xi}a_{r}\\
          imc^22\sqrt{\xi}a_{r}^{\dagger} & -mc^{2}
\end{pmatrix}.
\end{equation}

In this context, we introduce the right- and left-handed chirality
quantum basis.
\begin{align}
  \label{eq:chiral_basis}
  \ket{n_{r}} = \frac{1}{\sqrt{n_{r}!}}(a_{r}^{\dagger})^{n_{r}}\ket{0},
  \quad
  \ket{n_{l}} = \frac{1}{\sqrt{n_{l}!}}(a_{l}^{\dagger})^{n_{l}}\ket{0}.
\end{align}
The states $\ket{n_{r}}$ ($\ket{n_{l}}$) are the eigenstates
of the right (left)-handed number operators
$N_r=a_r^\dagger a_r$ ($N_l=a_l^\dagger a_l$).

The eigenvalues of the $(2{+}1)$ DO are \cite{UHDRE2022}
\begin{equation} \label{eq:energy_2+1}
    E_{2}^{\pm}=\pm mc^{2}\sqrt{4\xi(1+n_{r})+1}.
\end{equation}
By comparing the energies in Eqs. \eqref{eq:energy_1+1} and
\eqref{eq:energy_2+1}, we observe that they are similar, differing in
the constant factor multiplying $\xi$, and in the nature of the quanta
associated with the number operator.
Employing the ladder operators, Eq. \eqref{eq:ladder}, we can further
rewrite the $(2{+}1)$ DO Hamiltonian as
\begin{equation}
  \label{eq:do_2_map}
  H_2=  \hbar (g_2 a_{r}^{\dagger}\sigma_{-}+ g_2^* a_{r}\sigma_{+})
  +mc^{2}\sigma_{z},
\end{equation}
where $g_2$ is the spin-orbit coupling, given by
\begin{equation}
  \label{eq:coup_2}
  g_2
  = \frac{ imc^2 \sqrt{4\xi}}{\hbar}
  =\frac{ic}{\hbar}\sqrt{4 m\omega\hbar}.
\end{equation}
Thus, we again obtain a JC interaction, where the spin is now coupled to
the orbital angular momentum degree of freedom, specifically to the
right-handed chirality.
Had we chosen the negative sign, we would have instead obtained
left-handed chirality, accompanied by an AJC interaction
\cite{BERMUDEZ2007}.
Using the protocol established in Ref. \cite{UHDRE2022}, it is possible
to perform the mapping while simultaneously accounting for both
chiralities and interactions.

The difference regarding the $(1{+}1)$ case, Eq. \eqref{eq:do_1_map}, is
the scale of the coupling parameter $g_2$.
Additionally, as in the $(1{+}1)$ case, we can take a massless limit and
obtain a $(2{+}1)$ Weyl oscillator.
Recently, an analogue scenario was implemented in an experimental setup
\cite{Jiang2022}.

In this case, the conserved quantity reads
\begin{equation}
  \label{eq:conserved_2}
    I_2 = (N_r - N_l) + \frac{1}{2} \sigma_z.
\end{equation}
Physically, this conservation law corresponds to the total angular
momentum along the $z$-axis, $J_z$, given by
\begin{equation} \label{eq:total}
    J_z = L_z + S_z,
\end{equation}
where
\begin{align}
    L_z &= \hbar (N_r - N_l), \\
    S_z &= \frac{\hbar}{2} \sigma_z,
\end{align}
are the orbital and spin angular momentum operators, respectively.
Equation \eqref{eq:conserved_2} (or equivalently Eq. \eqref{eq:total})
generates a $U(1)$ symmetry in the $(2{+}1)$-dimensional case, in a
manner analogous to the role of $I_1$ in the $(1{+}1)$-dimensional
scenario.

\begin{figure}
  \centering
  \includegraphics[width=0.7\columnwidth]{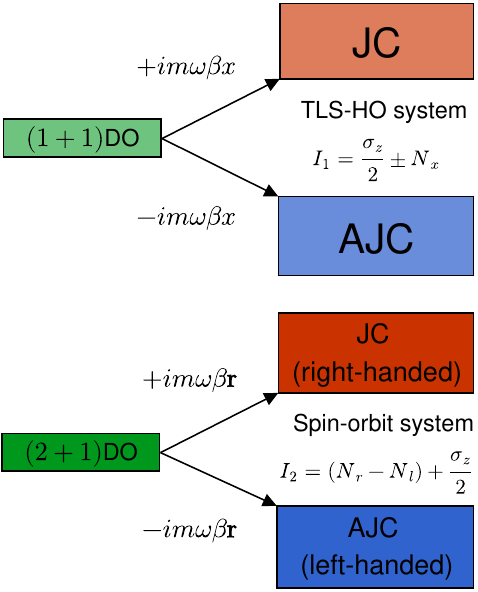}
  \caption{
    We summarize the different coupling signs and dimensions ($(1{+}1)$
    or $(2{+}1)$), along with their respective optical mappings.
    In the $(1{+}1)$ dimensions case, the system corresponds to the
    interaction of a TLS with an HO, yielding JC interactions for
    positive coupling and AJC interactions for negative coupling.
    The conserved quantity is symbolized by $I_1$. In the $(2{+}1)$
    case, the system exhibits spin–orbit coupling, with JC interactions
    associated with right-handed chirality for positive coupling, and
    AJC interactions associated with left-handed chirality for negative
    coupling.
    The conserved quantity is symbolized by $I_2$.
  }
  \label{fig:scheme}
\end{figure}

The mapping in Eq. \eqref{eq:do_2_map} was carried out exactly by
Bermudez \etal, who also proposed an ion-trap experiment based
on it.
In a subsequent work \cite{Bermudez2008a}, they studied the quantum
phase transition by considering the DO and a external magnetic field
within the Dirac equation, employing both JC and AJC mappings.
In that work, it was shown that the DO interaction can be interpreted as
analogous to the coupling of a fermion with a homogeneous transverse
electromagnetic field \cite{Castro2020}.
In Fig. \ref{fig:scheme}, we present a scheme for the mapping
considering the $(1{+}1)$ and $(2{+}1)$ dimensions cases, and both signs
in the coupling.

In conclusion, the $(1{+}1)$ and $(2{+}1)$ cases correspond to the same
mathematical problem.
They differ in the interpretation of the fermionic and bosonic degrees
of freedom, as well as in the scaling of the energy and coupling
parameters.
In the study of the dynamics, we adopt the $(2{+}1)$ DO for our
extension, presented in the next section, as it offers a more general
framework.
Nevertheless, the modifications we introduce can be readily applied to
the simpler $(1{+}1)$ case.
In this context, we employ $g_2=ig$ from now on, for a shorthand
notation.

\section{Time-dependent frequency}
\label{sec:time_dep}

Our extension consists of considering a time-dependent frequency, such
that $\omega \to \omega (t)$.
Naturally, considering Eqs. \eqref{eq:relativisitic_par} and
\eqref{eq:coup_2}, this leads to $\xi\to \xi(t)$ and $g \to g(t)$.
Therefore, we can observe the system transitioning from a relativistic
regime to a less relativistic one during its time evolution.
Furthermore, this categorizes our system as a time-dependent Dirac
equation problem \cite{Landim2000,DeCastro2003,Zhang2007}, a class that
occasionally admits analytical solutions.
One possible approach in this context is the Lewis–Riesenfeld invariant
method \cite{Lewis1969,Landim2000,DeCastro2003}, which allows the
determination of wave functions from a dynamical invariant operator.
In the case of a non-commutative phase space, this technique has been
employed for the Dirac oscillator with a time-dependent frequency
\cite{Sobhani2015}.
In contrast, our work follows a different approach: we study the system
on the basis of spin and angular momentum, focusing on the analysis of
entanglement and the expectation values of angular momentum observables.

Although we could consider a time-dependent mass as well
\cite{Zhang2007}, we restrict our focus to the case of time-dependent
frequency, thinking of a map with the TDJC model with a time-dependent
coupling parameter
\cite{Prants1992,Schlicher1989,Bartzis1992,Wilkens1992,
Joshi1993,Prants1997,Fang1998,Dasgupta1999}.
From the perspective of cavity quantum electrodynamics, this extension
could account for variations in the spatial profile of the cavity mode
experienced by an atom traversing the cavity
\cite{Haroche2006,Larson2021}.
That being the case, the time-dependent frequency may assist in
simulations of the DO in optical cavities.
An alternative approach to a time-dependent DO Hamiltonian was done by
Wang \cite{Wang2012}, considering a time-dependent laser pulse and
time-dependent perturbation theory.

In the computations that follow, we consider time-dependent quantities
in the form of
\begin{align}
    \omega(t) = {} &\omega_0 f^2(t), \\
    \xi(t) =  {} & \frac{\hbar \omega_0}{mc^2}f^2(t) = \xi_0 f^2(t), \\
    g(t) = {} & \frac{ 2mc^2 \sqrt{\xi(t)}}{\hbar} = g_0 f(t),
\end{align}
where $\xi_0=\hbar \omega_0/mc^2$ and $g_0=2c/\hbar \sqrt{ m\omega_0 \hbar}$.
We add that $g_0$ can be used as a control parameter to analyze different relativistic regimes, being proportional to $\xi_0$.
If we take the Weyl oscillator limit, we have
$g_0\to\mbox{const.}\neq 0$ \cite{Franco-Villafane2013}.

We shall work with the Hamiltonian in the form of
\begin{equation}
  \label{eq:do_2_map_time_dep}
  H_2(t)= i   \hbar g(t)(a_{r}^{\dagger}\sigma_{-}
  - a_{r}\sigma_{+})+mc^{2}\sigma_{z}.
\end{equation}
Notably, even within our generalization, the invariant quantities,
Eqs. \eqref{eq:conserved} and \eqref{eq:conserved_2}, continue to be
conserved.

In this context, since our Hamiltonian is now time-dependent, the
eigenvalues are no longer constant in time.
However, they are easily generalized with the proper modulation in the
relativistic parameter, i.e.,
\begin{equation}
  \label{eq:energy_2+1_time_dep}
    E_{2}^{\pm}(t)=\pm mc^{2}\sqrt{4\xi(t)(1+n_{r})+1}.
\end{equation}
Due to the explicit time dependence of the Hamiltonian, the
corresponding eigenvalues are time-dependent quantities defining an
instantaneous spectrum, rather than stationary energy levels
\cite{SAKURAI2020}.

Furthermore, we may switch to the interaction picture to eliminate the
free evolution term in Eq. \eqref{eq:do_2_map_time_dep} and analyze the
dynamics.
In this manner, we have the interaction Hamiltonian $V(t)$, given by
  \begin{equation}
    \label{eq:do_2_map_time_dep_int}
  V(t)=  i \hbar g(t) \left(a_{r}^{\dagger}\sigma_{-}e^{-2i\theta t} - a_{r}\sigma_{+}e^{2i\theta t} \right),
\end{equation}
where we use $\theta=mc^2/\hbar$.
A general state associated with the Hamiltonian
\eqref{eq:do_2_map_time_dep} has the form of
\begin{equation} \label{eq:state}
  \ket{\Psi(t)} = \sum_{n_r=0}^{\infty}
  \left[
    a_{n_r}(t)\ket{\uparrow,n_r} +
    b_{n_r+1}(t)\ket{\downarrow,n_r+1}
  \right],
\end{equation}
where $\ket{\uparrow}$ and $\ket{\downarrow}$ symbolize the Pauli
spinors \cite{BERMUDEZ2007}.
The ground state $\ket{\downarrow,0}$ is disregarded in
Eq. \eqref{eq:state},
as it will not be part of our initial state.
From the conservation of total angular momentum \eqref{eq:conserved_2},
it follows that the dynamics is confined to a $2{\times}2$ subspace.
Thus, from Schr\"odinger equation
\begin{equation}
    i\hbar \frac{d}{dt} |\Psi(t)\rangle = V(t)|\Psi(t)\rangle,
\end{equation}
applied to the state in Eq. \eqref{eq:state}, we obtain the set of
coupled differential equations
\begin{align}
  \dot{a}_{n_r}(t) = {} & -g(t)b_{n_r+1}(t)e^{i2\theta t}\sqrt{n_r+1},
   \\
  \dot{b}_{n_r+1}(t) = {} & g(t)a_{n_r}(t)e^{-i2\theta t}\sqrt{n_r+1}.
\end{align}
By decoupling the equations, we obtain
\begin{align}
  \label{eq:edos}
    \ddot{a}_{n_r}(t)-\left[\frac{\dot{g}(t)}{g(t)}+2\theta
  i\right]\dot{a}_{n_r}(t)+g^{2}(t)(n_r+1)a_{n_r}(t)  = {} & 0, \\
    \ddot{b}_{n_r+1}(t)-\left[\frac{\dot{g}(t)}{g(t)}-2\theta
  i\right]\dot{b}_{n_r+1}(t)+g^{2}(t)(n_r+1)b_{n_r+1}(t) = {} & 0.
\end{align}
These differential equations generally do not admit analytical solutions
for arbitrary time-dependent coupling.
Notable exceptions include the Rosen–Zener \cite{Rosen1932,Dasgupta1999}
and Nikitin \cite{Nikitin1984,Prants1992} models.
The latter will be examined in the context of the DO in subsection
\ref{subsec:exp}.

In addition to the expectation values of the angular momentum
observables, we also investigate spin-orbit entanglement.
To quantify this, we first obtain the reduced density matrix of the spin
subsystem by tracing out the orbital angular momentum
\begin{equation}
  \hat{\rho}_S (t)=\Tr_L [\hat{\rho}(t)] =
  \sum_{n_r=0}^{\infty}\langle n_r|\hat{\rho}(t) |n_r\rangle.
\end{equation}
We then employ the entanglement entropy \cite{Horodecki2009}, defined as
the von Neumann entropy \cite{VonNeumann1927b} of the subsystem.
For the spin subsystem, it is given by
\begin{equation} \label{eq:vne}
     \mathfrak{E}(t) = -\sum_i \mu_i(t) \log_2{\mu_i(t)},
\end{equation}
where $\mu_i(t)$ denotes the eigenvalues of $\hat{\rho}_S(t)$.

As the initial condition, we set $b_{n_r+1}(0)=0$, which yields the
initial state
\begin{equation}
  \label{eq:init_state}
  \ket{\Psi(0)} = \sum_{n_r=0}^{\infty}  a_{n_r}(0)\ket{\uparrow,n_r},
\end{equation}
with $a_{n_r}(0)$ characterizing the quanta distribution in orbital
angular momentum at time $t=0$.
For an initial number state, the coefficients are given by $ a_{n_r}(0)=\delta_{n_r,m_r}$, whereas for a circular coherent state \cite{BERMUDEZ2007}, they take the form
\begin{equation}
  a_{n_r}(0) =
  e^{-\frac{|\alpha_r|^2}{2}} \frac{\alpha_r^{n_r}}{\sqrt{n_r!}},
\end{equation}
where $|\alpha_r|^2=\expval{N_r(0)}$.
From an optical perspective, coherent states play a central role in
cavity quantum electrodynamics.
They are characterized by a Poisson distribution and are particularly
relevant due to their greater experimental accessibility
\cite{Haroche2006,Zhang2024}.
In ion-trap experiments, both the number and coherent states can be
achieved, with the ion quanta of vibration \cite{BERMUDEZ2007}.

We symbolize the expectation values of the spin, orbital angular
momentum, and total angular momentum for an initial number state in the
orbital angular momentum subsystem as $\expval{S_z^{n_r}(t)}$,
$\expval{L_z^{n_r}(t)}$ and $\expval{J_z^{n_r}(t)}$, respectively.
The expectation values of the same observables for an initial coherent
state read as follows:
\begin{align}
  \expval{S_z(t)} = {}
  &
    \sum_{n_r=0}^\infty \frac{e^{-|\alpha_r|^2}|\alpha_r|^2}{n_r!}
    \expval{S_z^{n_r}(t)},
    \label{eq:gen_1_coh} \\
  \expval{L_z(t)} = {}
  &
    \sum_{n_r=0}^\infty \frac{e^{-|\alpha_r|^2}|\alpha_r|^2}{n_r!}
    \expval{L_z^{n_r}(t)},
    \label{eq:gen_2_coh} \\
  \expval{J_z(t)\rangle}
  = {}
  &
    \sum_{n_r=0}^\infty \frac{e^{-|\alpha_r|^2}|\alpha_r|^2}{n_r!}
    \expval{J_z^{n_r}(t)}.
    \label{eq:gen_3_coh}
\end{align}

Three representative modulations are considered: constant, exponential,
and sinusoidal. 
The constant case serves as a reference to highlight the contrast when a time-dependent frequency is introduced. 
The exponential modulation encompasses two key features of TDJC model
coupling: it represents temporal dependencies responsible for dynamic
acceleration or deceleration and admits analytical solutions in the
off-resonant regime. 
In this same regime, the sinusoidal modulation is associated with
aperiodic evolution.

\subsection{Constant frequency}
\label{sec:const}
\begin{figure}
  \centering
  \includegraphics[width=\columnwidth]{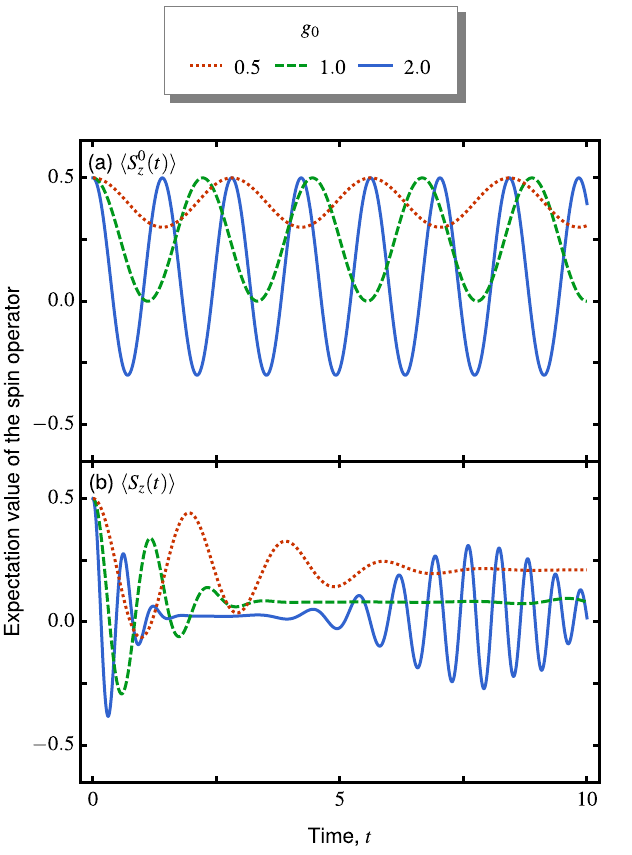}
  \caption{
  Expectation value of the spin observable, using units such that
  $m=c=\hbar=1$.
  We analyze different relativistic regimes, employing: $g_0=0.5$
  (dotted red line), $g_0=1.0$ (dashed green line) and $g_0=2.0$
  (solid blue line).
  In (a), we consider an initial number state with $n_r=0$, while in
  (b), we assume the initial coherent state with $|\alpha_r|^2=5$.
  }
  \label{fig:const_spin}
\end{figure}
First, we consider the standard scenario, where
$f(t)=\mbox{const.} \neq 0$ and  $g(t)=g_0 \neq 0$.
Without loss of generality, we set the constant function $f(t)=1$.
In this case, the energy eigenvalues are given by
Eq. \eqref{eq:energy_2+1}, with the change $\xi\to\xi_0$.
The solutions of the ODEs, Eq. \eqref{eq:edos}, are given by
\begin{align}
  a_{n_r}(t) =  {}
  &
    e^{i \theta t} a_{n_r}(0) \left[ \cos(\Theta_{n_r} t)
    - \frac{i \theta}{\Theta_{n_r}} \sin(\Theta_{n_r} t) \right], \\
  b_{n_r+1}(t) = {}
  &
    e^{-i \theta t} a_{n_r}(0)
    \frac{g_0 \sqrt{n_r+1}}{\Theta_{n_r}} \sin(\Theta_{n_r} t),
\end{align}
where we have introduced the relativistic Rabi frequency,
\begin{equation}
  \Theta_{n_r}=\sqrt{\theta ^2+ g_0^2 (n_r+1)}.
\end{equation}
With the probability amplitudes at hand, we can compute the expectation
values of the observables.
For an initial number state, we obtain
%\revision{
%\begin{align}
%  \expval{S_z^{n_r}(t)}  = {}
%  &
%    \frac{\hbar}{2} \left[\frac{ \theta ^2+ g_0^2 (n_r+1)
%    \cos \left(2 \Theta_{n_r} t \right)}{\Theta_{n_r}^2} \right], \\
%  \expval{L_z^{n_r}(t)}  =  {}
%  &
%   \frac{\hbar}{2} \left[\frac{- g_0^2 (n_r+1) \cos \left(2 \Theta_{n_r} t \right)+g_0^2 (n_r+1) (2 n_r+1)+2\theta ^2 n_r}{ % \Theta_{n_r}^2}\right], \\
%\expval{J_z^{n_r}(t)} = {}
%&
%\frac{\hbar}{2}(1 + 2n_r).
%\end{align}}
\begin{align}
  \expval{S_z^{n_r}(t)}  = {}
  &
    \frac{\hbar}{2\Theta_{n_r}^2} \Big[ \theta ^2+ g_0^2 (n_r+1) \cos \left(2 \Theta_{n_r} t \right) \Big], \\
    \expval{L_z^{n_r}(t)}  =  {}
  &
   \frac{\hbar}{2 \Theta_{n_r}^2} \Big[- g_0^2 (n_r+1) \cos \left(2 \Theta_{n_r} t \right)  +g_0^2 (n_r+1) \nonumber \\ 
  & \times (2 n_r+1) +2\theta ^2 n_r\Big], \\
  \expval{J_z^{n_r}(t)} = {}
  &
    \frac{\hbar}{2}(1 + 2n_r).
\end{align}

\begin{figure}
  \centering
  \includegraphics[width=\columnwidth]{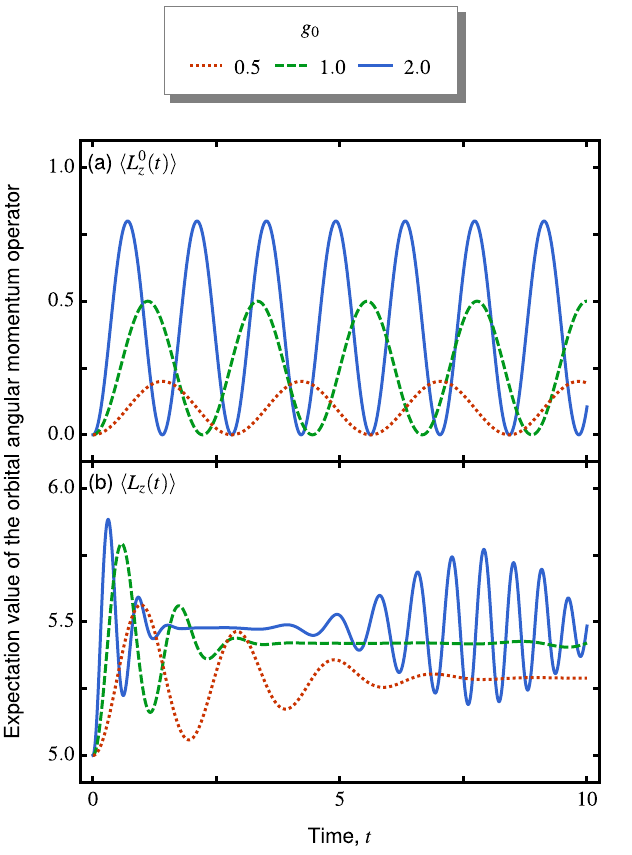}
  \caption{
   Expectation value of the orbital angular momentum observable, using
   units such that $m=c=\hbar=1$.
  We analyze different relativistic regimes, employing: $g_0=0.5$
  (dotted red line), $g_0=1.0$ (dashed green line) and $g_0=2.0$ (solid
  blue line).
  In (a), we consider an initial number state with $n_r=0$, while in
  (b), we assume the initial coherent state with $|\alpha_r|^2=5$. 
  }
  \label{fig:const_l}
\end{figure}

The behavior of the observables for an initial coherent state follows
from Eqs. \eqref{eq:gen_1_coh}, \eqref{eq:gen_2_coh} and
\eqref{eq:gen_3_coh}.
For the total angular momentum, we have
\begin{equation}
  \expval{J_z(t)}  = \frac{\hbar}{2}\left(1 + 2|\alpha_r|^2\right).
\end{equation}
To display the time evolution of the expectation value of the
observables, we employ units such that $m=c=\hbar=1$.
The expectation values for the spin operator are displayed in
Fig. \ref{fig:const_spin} for different relativistic regimes, employing
three values for $g_0$: $g_0=0.5$ (dotted red line), $g_0=1.0$ (dashed
green line), and $g_0=2.0$ (solid blue line).
When the initial state of the orbital angular momentum is a number state, specifically the number $n_r=0$,
the system exhibits periodic oscillations confined within the interval
$\pm 1/2$, as illustrated in Fig. \ref{fig:const_spin}(a).
These oscillations characterize the spin \emph{Zitterbewegung},
a purely relativistic phenomenon arising from the interference between
positive- and negative- energy components.
We observe that for smaller values of $g_0$, the system exhibits reduced
oscillation amplitudes and a slower time evolution. 
The parameter $g_0$, which quantifies the spin-orbit coupling strength,
is also proportional to the relativistic parameter $\zeta$
(Eq. \eqref{eq:coup_2}).
Thus, the dynamical dependence of the observables on $g_0$ can be
explained by the system's proximity to the relativistic and
nonrelativistic limits.
The oscillations amplitude does not attain its maximal negative value
($-1/2$) due to the contribution of the rest mass term.
On the other hand, when the initial state of the orbital angular
momentum is taken to be a coherent state with $|\alpha_r|^2=5$, as shown
in Fig. \ref{fig:const_spin}(b), the system exhibits an initial collapse
of the oscillations, followed by a quiescent period and a subsequent
revival.
This behavior arises from the underlying Poissonian distribution of
phonons, which leads to a superposition of terms that undergo dephasing
and rephasing over time.
This behavior extends to the orbital angular momentum observable,
Figs. \ref{fig:const_l}(a) and (b), where the oscillations exhibit an
inverted pattern relative to the spin, ensuring that the total angular
momentum remains conserved.

Experimentally probing relativistic fermions poses significant
challenges due to the high energy scales involved \cite{BERMUDEZ2007}.
As an alternative, their dynamics can be simulated in controllable
systems such as trapped ions.
In this setting, \emph{Zitterbewegung} has been observed for a massive
free particle \cite{Gerritsma2010}.
The influence of an external magnetic field has also been explored:
\emph{Zitterbewegung} was measured in systems simulating both the Dirac
equation \cite{Rusin2010} and the Weyl equation \cite{Jiang2022}.

\begin{figure}
  \centering
  \includegraphics[width=1\columnwidth]{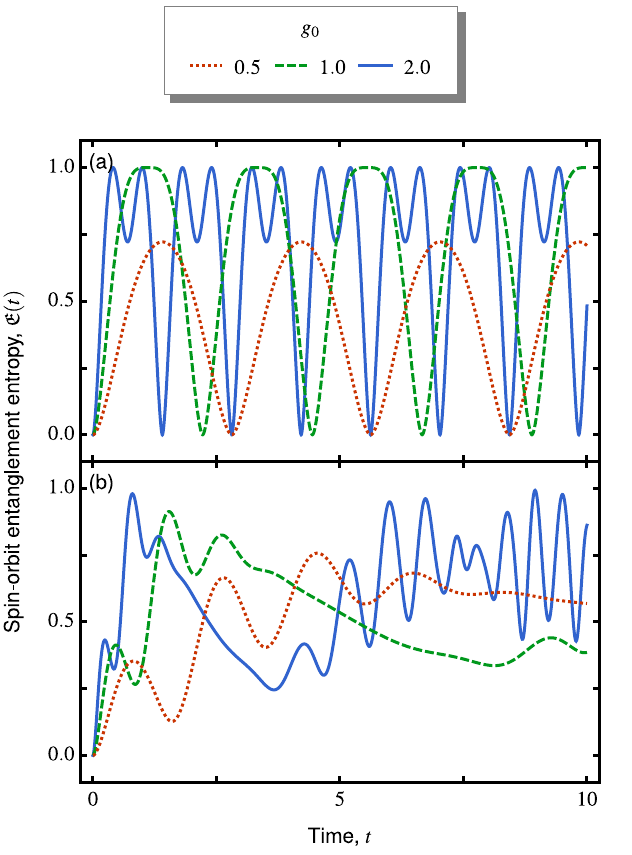}
  \caption{
    Time evolution of the spin-orbit entanglement, using units such that
    $m=c=\hbar=1$.
    We analyze different relativistic regimes, employing: $g_0=0.5$
    (dotted red line), $g_0=1.0$ (dashed green line) and $g_0=2.0$ (solid
    blue line).
    In (a), we consider an initial number state with $n_r=0$, while in
    (b),   we assume the initial coherent state with $|\alpha_r|^2=5$.
  }
  \label{fig:const_vne}
\end{figure}

The spin-orbit entanglement, defined in Eq. \eqref{eq:vne}, is shown in
Fig. \ref{fig:const_vne}, using the parameters previously introduced.
When the system is initially prepared in a number state, the reduced
subsystem undergoes periodic oscillations between mixed and pure states,
resulting in alternating regions of high and low entanglement with a
fixed periodicity.
The entanglement reaches its maximum when the probabilities are equal,
which occurs at the instants when $\langle S_z(t) \rangle = 0$, as shown
in Fig. \ref{fig:const_spin}.
In contrast, when the initial state is a coherent state, the
off-diagonal elements of the spin reduced density matrix are non-zero,
leading to a different evolution.
After the initial oscillations, during the collapse time-interval,
entanglement reaches a minimum -- this is especially clear for the case
where $g_0=2.0$.
In this instant, the state evolves into an almost separable state.
In the language of the JC model, this behavior was analyzed in Ref. \cite{Phoenix1991}.

\subsection{Exponential frequency}
\label{subsec:exp}

Now, we consider a time-dependent modulation in the form
\begin{equation}
  \label{eq:exp_mod}
    f(t)= e^{\zeta t},
\end{equation}
which allows for analytical solutions of the differential equations in
Eq. \eqref{eq:edos}.
At first glance, the form we have chosen for this modulation appears to
diverge.
However, we constrain $\zeta$ to negative values in the  graphical
representations of the quantities studied.
Alternatively, in the framework of $\zeta>0$, one could choose a small
intensity for it or consider a shorter time frame.
This is the relativistic version of the Nikitin model \cite{Nikitin1984},
and was studied in the JC perspective in Ref. \cite{Prants1992}.
This may account for transient effects in an optical cavity, such as
switching on and off.

\begin{figure}
  \centering
  \includegraphics[width=1\columnwidth]{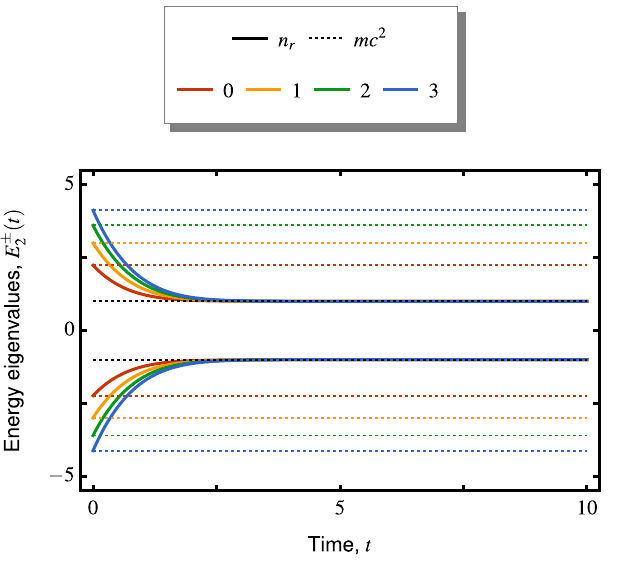}
  \caption{
    Time evolution of the energy eigenvalues considering the exponential
    modulation, Eq. \eqref{eq:energy_2+1_exp}, using units such that
    $m=c=\hbar=1$ and $\xi_0=1$, and considering different quanta:
    $n_r=0$ (solid red line), $n_r=1$ (solid yellow line), $n_r=2$ (solid
    green line), $n_r=3$ (solid blue line).
    The rest mass energies, $\pm mc^2$, are shown as dashed black lines,
    while the constant energy eigenvalues from Eq. \eqref{eq:energy_2+1}
    are represented by dashed colored lines, matching the colors of
    their corresponding time-dependent counterparts.
  }
  \label{fig:exp_energy}
\end{figure}

In this scenario, the instantaneous energy eigenvalues are given by
\begin{equation}
  \label{eq:energy_2+1_exp}
    E_{2}^{\pm}(t)=\pm mc^{2}\sqrt{4\xi_0\exp(2\zeta t)(1+n_{r})+1},
\end{equation}
where, for $\zeta<0$, we obtain asymptotic values, obtaining the minimum
energy $\pm mc^2$,  when $t \to \infty$.
With this modulation, we can see the system transitioning from a
relativistic regime to a nonrelativistic one.
At long times, this implies the uncoupling of the spin and angular
degrees of freedom.
Using the previously established unit system, we present the evolution
of Eq. \eqref{eq:energy_2+1_exp} in Fig. \ref{fig:exp_energy}, fixing
$g_0=2.0$ and $\zeta=-1$.
Naturally, larger quanta imply in larger absolute values of initial
energy, and a slower decay to the rest mass.

Returning to the differential equations, Eq. \eqref{eq:edos}, we employ
Eq. \eqref{eq:exp_mod}, obtaining
\begin{multline}
  \label{eq:edos_tau}
    \frac{d^{2}{a}_{n_r}(t)}{dt^{2}}-\left(\zeta+ 2\theta i \right)\frac{d{a}_{n_r}(t)}{dt}
    % &+ \nonumber
    \\
    +g_0^2\exp(2 \zeta t)(n_r+1)a_{n_r}(t) =0,
  \end{multline}
  \begin{multline}
    \frac{d^{2}{b}_{n_r+1}(t)}{dt^{2}}-\left(\zeta-2\theta i \right)\frac{d{b}_{n_r+1}(t) }{dt}
    % &+\nonumber
    \\
    +g_0^2\exp(2 \zeta t)(n_r+1)b_{n_r+1}(t)  =0.
\end{multline}
The solution to the equations above is
\begin{align}
  a_{n_r}(t) = {}
  &
    a_{n_r}(0) e^{\frac{1}{2} t (\zeta +2 \theta  i)} Z^a_\eta(z_{n_r}(t)), \\
  b_{n_r+1}(t) = {}
  &
   -a_{n_r}(0)e^{\frac{1}{2} t (\zeta -2 \theta i)} Z^b_\eta(z_{n_r}(t)).
\end{align}
We have employed, for a shorthand notation, the following parameters
\begin{align}
  \eta = {}
  &
    \frac{1}{2}-\frac{\theta i}{\zeta}, \\
    % \beta&=\pm\frac{1}{2}(1-\frac{2\theta i}{\zeta}), \\
    z_{n_r}(t)  = {}
   &
     \frac{g_{0}\sqrt{n_r+1}\exp(\zeta t)}{\zeta},
\end{align}
and the functions
\begin{align}
 &  Z_\eta^a(z_{n_r}(t)) = 
    z_{n_r}(0) \frac{\pi}{2}
    \sech \left(\frac{\pi  \theta}{ \zeta }\right) \nonumber\\
& \times
    \left[ J_{\eta}\left(z_{n_r}(0)\right) J_{-\eta+1}\left(z_{n_r}(t)\right)
    +J_{-\eta}\left(z_{n_r}(0)\right) J_{\eta-1}\left(z_{n_r}(t)\right) \right] ,\\
&   Z_\eta^b(z_{n_r}(t))= {}
    z_{n_r}(0) \frac{\pi}{2}  \sech \left(\frac{\pi  \theta}{ \zeta }\right)  \nonumber  \\
&  \times \left[ J_{\eta}\left(z_{n_r}(0)\right) J_{-\eta}\left(z_{n_r}(t)\right)-J_{-\eta}\left(z_{n_r}(0)\right)     J_{\eta}\left(z_{n_r}(t)\right)\right].
\end{align}
The Bessel functions are symbolized by
$J_x(w(t))$\cite{Prants1992,Arfken2013}, and depend on $z_{n_r}(t)$ and
$\eta$.
We can take a limit to derive expressions for  the asymptotic values of
the probability amplitudes, considering $t\to\infty$ and $\zeta<0$:
\begin{align}
\!\!  \lim_{t\to\infty} a_{n_r}(t) = {}
  &
    a_{n_r}(0) \frac{\pi}{2^\eta} \text{sech}\left(\frac{\pi  \theta }{
    \zeta }\right) \frac{ (z_{n_r}(0))^{\eta }  J_{-\eta
    }(z_{n_r}(0))}{\Gamma (\eta )},
    \label{eq:asym_1} \\
\!\!  \lim_{t\to\infty} b_{n_r+1}(t) = {}
  &
    a_{n_r}(0) \frac{\pi}{2^{1-\eta}} \text{sech}\left(\frac{\pi  \theta}{ \zeta }\right) \frac{(z_{n_r}(0))^{1-\eta }  J_{\eta }(z_{n_r}(0))}{\Gamma (1-\eta )}, \label{eq:asym_2}
\end{align}
where $\Gamma(x)$ symbolizes the gamma function \cite{Arfken2013}.

For an initial number state in the orbital degree of freedom, the
expectation values of the observables are given by
\begin{align}
  \expval{S_z^{n_r}(t)}  = {}
  &
    \frac{\hbar}{2} e^{\zeta t}
    \left[|Z_\eta^a(z_{n_r}(t))|^2 - |Z_\eta^b(z_{n_r}(t))|^2 \right], \\
  \expval{ L_z^{n_r}(t)} = {}
  &
    \hbar e^{\zeta t}
    \left[n_r|Z_\eta^a(z_{n_r}(t))|^2 + (n_r+1)|Z_\eta^b(z_{n_r}(t))|^2 \right].
\end{align}
On the other hand, for an initial coherent state, the behavior follows
from the sum expressed in Eqs. \eqref{eq:gen_1_coh},
\eqref{eq:gen_2_coh} and \eqref{eq:gen_3_coh}.
The total angular momentum preserves the number of quanta, as in the
case of constant coupling, regardless of the initial state.

\begin{figure}
  \centering
  \includegraphics[width=1\columnwidth]{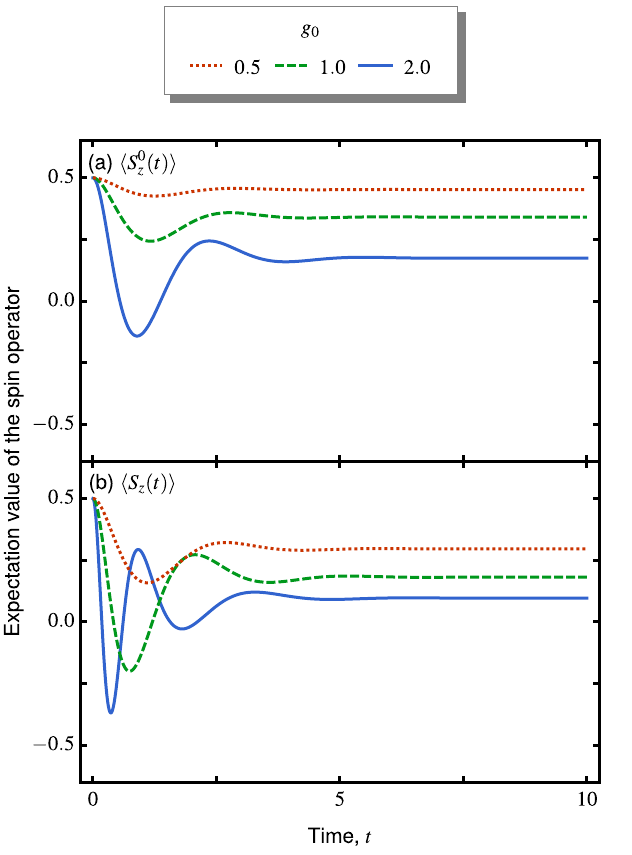}
  \caption{
    Expectation value of the spin observable, using units such that
    $m=c=\hbar=1$ and the exponential modulation, Eq. \eqref{eq:exp_mod},
    with $\zeta=-1$.
    We analyze different relativistic regimes, employing: $g_0=0.5$
    (dotted red line), $g_0=1.0$ (dashed green line) and $g_0=2.0$ (solid
    blue line).
    In (a) we consider an initial number state and $n_r=0$.
    In (b) we consider the initial coherent state with $|\alpha_r|^2=5$.
  }
  \label{fig:exp_spin}
\end{figure}

\begin{figure}
  \centering
  \includegraphics[width=1\columnwidth]{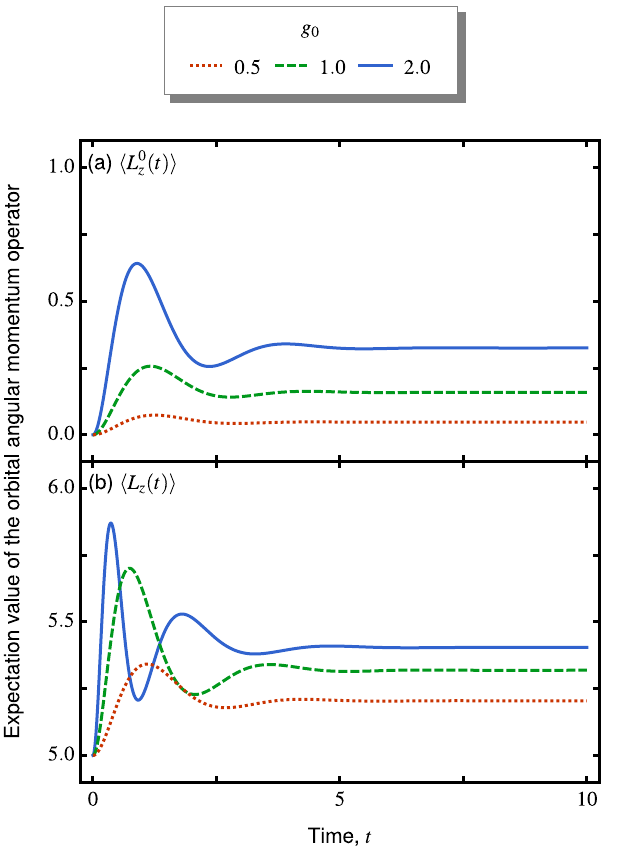}
  \caption{
    Expectation value of the orbital angular momentum observable, using
    units such that $m=c=\hbar=1$ and the exponential modulation,
    Eq. \eqref{eq:exp_mod}, with $\zeta=-1$.
    We analyze different relativistic regimes, employing: $g_0=0.5$
    (dotted red line), $g_0=1.0$ (dashed green line) and $g_0=2.0$ (solid
    blue line).
    In (a) we consider an initial number state and $n_r=0$.
    In (b) we consider the initial coherent state with $|\alpha_r|^2=5$.
  }
  \label{fig:exp_l}
\end{figure}

\begin{figure}
  \centering
  \includegraphics[width=1\columnwidth]{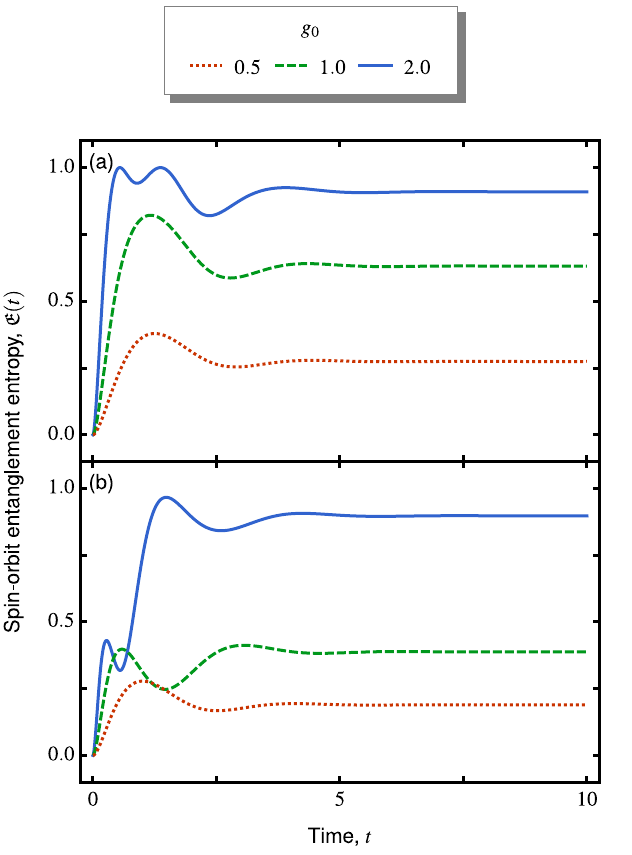}
  \caption{
    Time evolution of the spin-orbit entanglement, using units such that
    $m=c=\hbar=1$ and the exponential modulation,
    Eq. \eqref{eq:exp_mod}, with $\zeta=-1$.
    We analyze different relativistic regimes, employing: $g_0=0.5$
    (dotted red line), $g_0=1.0$ (dashed green line) and $g_0=2.0$
    (solid blue line).
    In (a) we consider an initial number state and $n_r=0$.
    In (b) we consider the initial coherent state with $|\alpha_r|^2=5$.
  }
  \label{fig:exp_vne}
\end{figure}

This asymptotic behavior of Eqs. \eqref{eq:asym_1} and \eqref{eq:asym_2}
manifest in all observables (see Figs. \ref{fig:exp_spin} and
\ref{fig:exp_l}).
Moreover, markedly different entanglement values were obtained across
the distinct relativistic regimes considered at the outset, as shown in
Fig. \ref{fig:exp_vne}.
Overall, smaller values of $g_0$ result in a faster decay of the
system’s dynamics, which, depending on the point at which the dynamics
vanish, impacts the resulting asymptotic spin-orbit entanglement.

The vanishing behavior of the exponential coupling for $\zeta<1$ can
also be modulated with a hyperbolic secant time dependence.
This form also permits analytical solutions in the off-resonance
scenario and models a physical picture similar to that of the
exponential modulation \cite{Dasgupta1999}. 
Conversely, choosing $\zeta>1$ would yield relativistic Rabi
oscillations with an increasingly larger frequency.
This behavior is qualitatively similar to, though faster than, that
produced by the linear coupling \cite{Joshi1993}, which is employed to
model a well-localized atom in a spatially varying mode
\cite{Larson2021}. 

\subsection{Sinusoidal frequency}
The semiclassical motion of an atom in a standing-wave cavity
\cite{Larson2021} leads to a TDJC Hamiltonian that includes a
trigonometric coupling term, which can be modulated by
\begin{equation}
  \label{eq:sin_mod}
  f(t)=\sin{(\zeta t)}.
\end{equation}
In the TDJC literature, this is one of the most studied scenarios
\cite{Schlicher1989,Bartzis1992,Wilkens1992,Fang1998}. 
We highlight the work of Prants and Kon'kov, who investigated the
off-resonant case, a situation that will be particularly useful in the
present work \cite{Prants1997}.

\begin{figure}
  \centering
  \includegraphics[width=1\columnwidth]{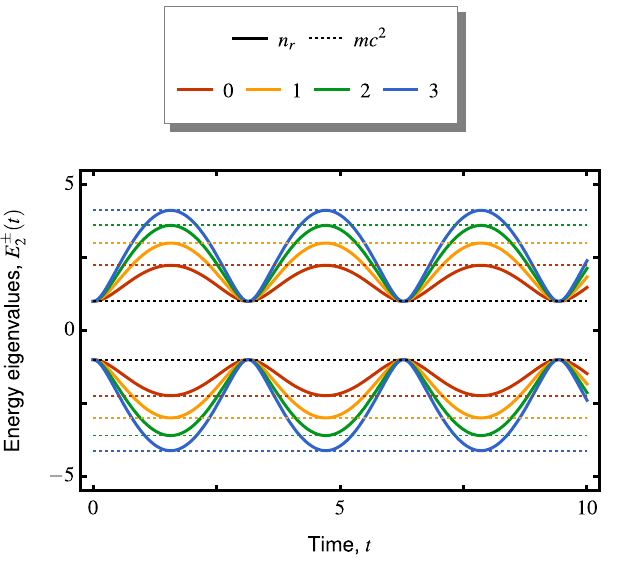}
  \caption{
    Time dependence of the instantaneous energy eigenvalues considering
    the sinusoidal modulation, Eq. \eqref{eq:sin_mod}, using units such
    that $m=c=\hbar=1$ and $\xi_0=1$, considering different quanta:
    $n_r=0$ (solid red line), $n_r=1$ (solid yellow line), $n_r=2$
    (solid green line), $n_r=3$ (solid blue line).
    The rest mass energies, $\pm mc^2$, are shown as dashed black lines,
    while the constant energy eigenvalues from Eq. \eqref{eq:energy_2+1}
    are represented by dashed colored lines, matching the colors of
    their corresponding time-dependent counterparts.
  }
  \label{fig:sin_energy}
\end{figure}

The instantaneous energy eigenvalues in this scenario are given by
\begin{equation}
  \label{eq:energy_2+1_sin}
  E_{2}^{\pm}(t)=\pm mc^{2}\sqrt{4\xi_0\sin^2(\zeta t)(1+n_{r})+1},
\end{equation}
and are displayed in Fig. \ref{fig:sin_energy}.
We observe oscillations between the maximum energy associated with a
given quantum number and the minimum value $mc^2$, with larger $n_r$
leading to more pronounced oscillations.
Effectively, the system transitions between less and more relativistic
regimes.

In this case, analytical solutions are not known without further
approximations.
One possible approach is to consider the aforementioned Weyl oscillator
limit \cite{Franco-Villafane2013},by taking $\theta \to 0$ and
$g_0 \to \text{const.} \neq 0$ (equivalently
$m \omega_0 \to \text{const.} \neq 0$).
This limit eliminates the rest mass, thereby reducing the dynamics to
that of a Weyl fermion \cite{Jiang2022} subject to a time-dependent
linear potential.

\begin{figure}
  \centering
  \includegraphics[width=1\columnwidth]{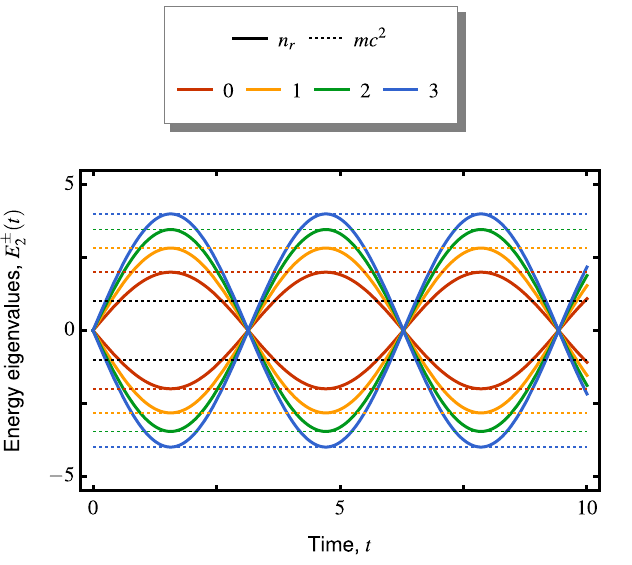}
  \caption{
    Time evolution of the instantaneous energy eigenvalues considering
    the sinusoidal modulation, Eq. \eqref{eq:sin_mod}, and the Weyl
    oscillator limit, using units such that $m=c=\hbar=1$ and
    $\omega_0=1$, considering different quanta: $n_r=0$ (solid red
    line), $n_r=1$ (solid yellow line), $n_r=2$ (solid green line),
    $n_r=3$ (solid blue line).
    The rest mass energies, $\pm mc^2$, are shown as dashed black lines,
    while the constant energy eigenvalues from Eq. \eqref{eq:energy_2+1}
    are represented by dashed colored lines, matching the colors of
    their corresponding time-dependent counterparts.
  }
  \label{fig:weyl_sin_energy}
\end{figure}

The instantaneous energy eigenvalues in this scenario are given by
\begin{equation}
  \label{eq:weyl_energy_2+1_exp}
    E_{2}^{\pm}(t)=\pm c\sqrt{4m\hbar\omega_0\sin^2(\zeta t)(1+n_{r})},
\end{equation}
and are displayed in Fig. \ref{fig:weyl_sin_energy}.
We observe a behavior similar to that in Fig. \ref{fig:sin_energy},
but now without the bound in the rest mass energy, resulting in
oscillations with larger amplitudes.

In the same limit, the differential equations become
\begin{align} \label{eq:edos_sin_weyl}
  \ddot{a}_{n_r}(t)-\frac{\dot{g}(t)}{g(t)}\dot{a}_{n_r}(t)
  +g^{2}(t)(n_r+1)a_{n_r}(t)  = {} & 0, \\
  \ddot{b}_{n_r+1}(t)-\frac{\dot{g}(t)}{g(t)}\dot{b}_{n_r+1}(t)
  +g^{2}(t)(n_r+1)b_{n_r+1}(t)  = {} & 0.
\end{align}
The solutions are
\begin{align}
  a_{n_r}(t)  = {}
  &
    a_{n_r}(0) \cos
    \Bigg( g_0\sqrt{n_r+1}\frac{  [\cos (\zeta  t)-1]}{\zeta }\Bigg), \\
  b_{n_r+1}(t) {} =
  &
    - a_{n_r}(0) \sin
    \Bigg( g_0\sqrt{n_r+1}\frac{  [\cos (\zeta  t)-1]}{\zeta }\Bigg).
\end{align}
In fact, in this limit, we can generalize the coefficients for an
arbitrary time-dependent modulation $g(t)$ \cite{Larson2021}:
\begin{align}
    a_{n_r}(t)  = {} &  a_{n_r}(0) \cos
    \Bigg(\sqrt{n_r+1} \int_0^t g(t') dt' \Bigg), \\
     b_{n_r+1}(t)  = {} & - i a_{n_r}(0) \sin
    \Bigg(\sqrt{n_r+1} \int_0^t g(t') dt \Bigg).
\end{align}

\begin{figure}
  \centering
  \includegraphics[width=1\columnwidth]{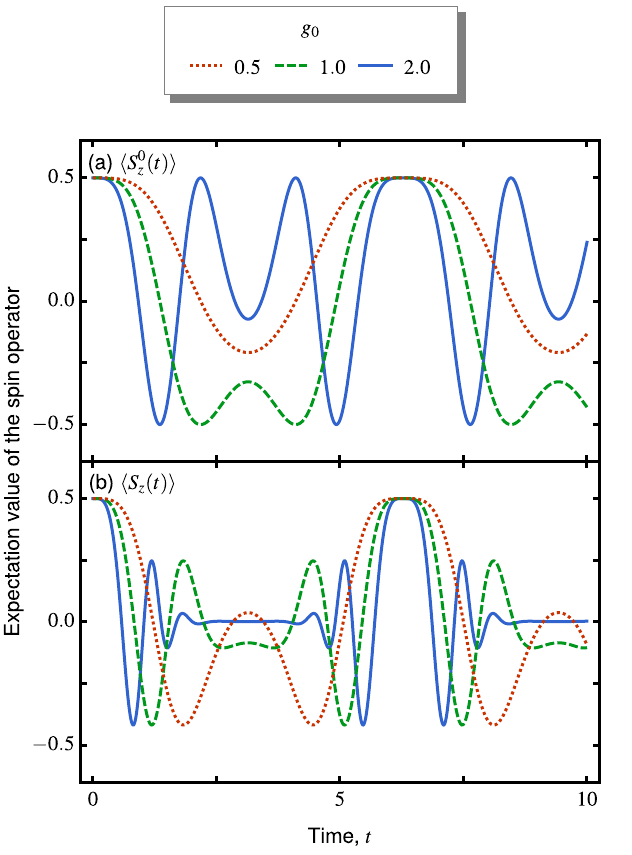}
  \caption{
    Expectation value of the spin observable in the Weyl oscillator
    limit, using units such that $m=c=\hbar=1$ and the sinusoidal
    modulation, Eq. \eqref{eq:sin_mod}, with $\zeta=1$.
    We analyze different relativistic regimes, employing: $g_0=0.5$
    (dotted red line), $g_0=1.0$ (dashed green line) and $g_0=2.0$
    (solid  blue line).
    In (a), we consider an initial number state with $n_r=0$, while in
    (b),  we assume the initial coherent state with $|\alpha_r|^2=5$.
  }
  \label{fig:weyl_sin_spin}
\end{figure}

\begin{figure}
  \centering
  \includegraphics[width=1\columnwidth]{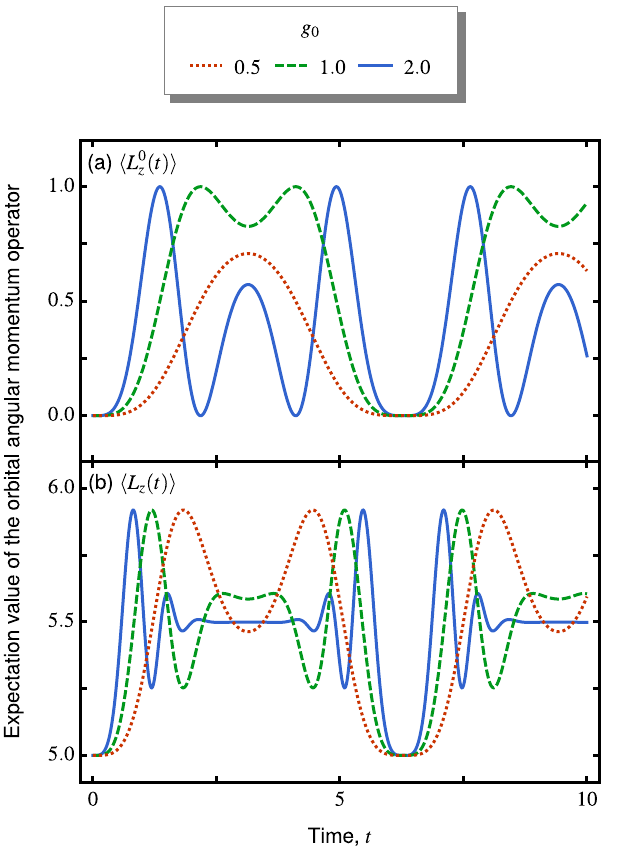}
  \caption{
  Expectation value of the orbital angular momentum observable in the
  Weyl oscillator limit, using units such that $m=c=\hbar=1$ and the
  sinusoidal modulation, Eq. \eqref{eq:sin_mod}, with $\zeta=1$.
  We analyze different relativistic regimes, employing: $g_0=0.5$
  (dotted red line), $g_0=1.0$ (dashed green line) and $g_0=2.0$ (solid
  blue line).
  In (a), we consider an initial number state with $n_r=0$, while in (b),
  we assume the initial coherent state with $|\alpha_r|^2=5$.
  }
  \label{fig:weyl_sin_l}
\end{figure}

\begin{figure}
  \centering
  \includegraphics[width=1\columnwidth]{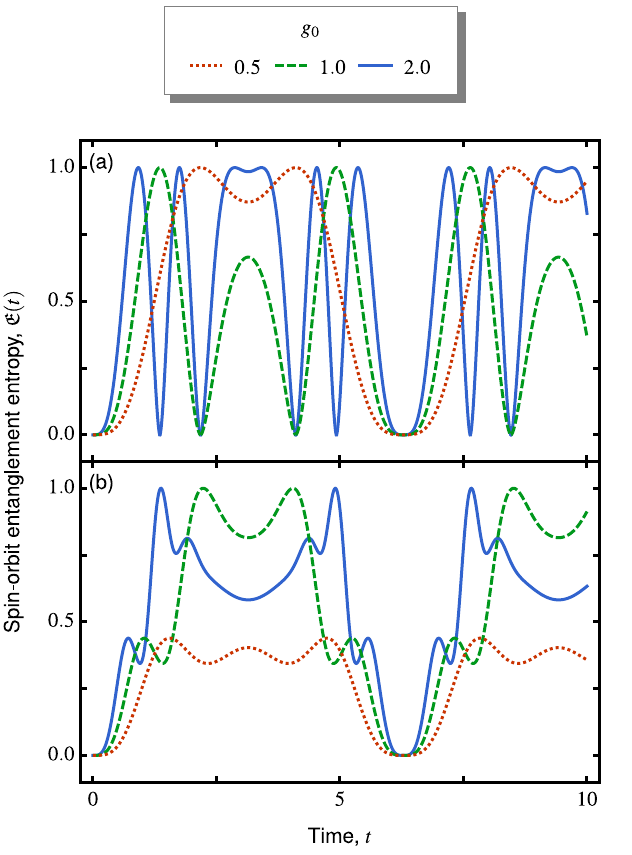}
  \caption{
    Time evolution of the spin-orbit entanglement in the Weyl
    oscillator limit, using units such that $m=c=\hbar=1$ and the
    sinusoidal modulation, Eq. \eqref{eq:sin_mod}, with $\zeta=1$.
  We analyze different relativistic regimes, employing: $g_0=0.5$
  (dotted red line), $g_0=1.0$ (dashed green line) and $g_0=2.0$ (solid
  blue line).
  In (a), we consider an initial number state with $n_r=0$, while in
  (b), we assume the initial coherent state with $|\alpha_r|^2=5$. 
  }
  \label{fig:weyl_sin_vne}
\end{figure}

From the probability amplitudes, the expectation values of the
observables follow.
For an initial number state, we have
\begin{align}
\expval{S_z^{n_r}(t)} = {}
  &
 \frac{\hbar}{2} \cos \Bigg(2 g_0 \sqrt{n_r+1}\frac{ \left[ \cos (\zeta  t)-1 \right]}{\zeta }\Bigg), \\
\expval{L_z^{n_r}(t)} = {}
  &
    \frac{\hbar}{2}\left[(1 + 2 n_r)
    - \cos \Bigg(2 g_0 \sqrt{n_r+1}\frac{ \left[ \cos (\zeta  t)-1 \right]}{\zeta }\Bigg)\right].
\end{align}
The behavior of these observables is characterized by periodicity, as
presented in Figs. \ref{fig:weyl_sin_spin} and \ref{fig:weyl_sin_l}.
Even when the initial phonon state is coherent, the collapse and revival
structure disappears, giving place to a purely periodic pattern.
Fang \cite{Fang1998} explains that this occurs because time is
dynamically rescaled by a trigonometric function.
This explanation can be extended to the spin-orbit entanglement, which
also exhibits the same periodicity, as shown in
Fig. \ref{fig:weyl_sin_vne}.

\begin{figure}
  \centering
  \includegraphics[width=1\columnwidth]{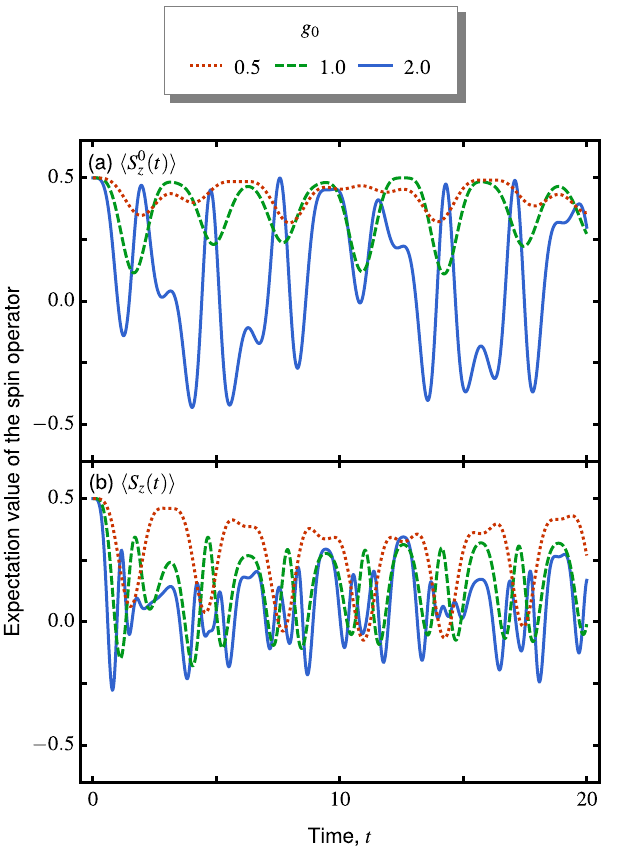}
  \caption{
    Numerical expectation value of the spin observable, using units such
    that $m=c=\hbar=1$ and the sinusoidal modulation, Eq. \eqref{eq:sin_mod},
    with $\zeta=1$.
    We analyze different relativistic regimes, employing: $g_0=0.5$
    (dotted red line), $g_0=1.0$ (dashed green line) and $g_0=2.0$
    (solid blue line).
    In (a), we consider an initial number state with $n_r=0$, while in (b),   we assume the initial coherent state with $|\alpha_r|^2=5$.
  }
  \label{fig:sin_spin}
\end{figure}

\begin{figure}
  \centering
  \includegraphics[width=1\columnwidth]{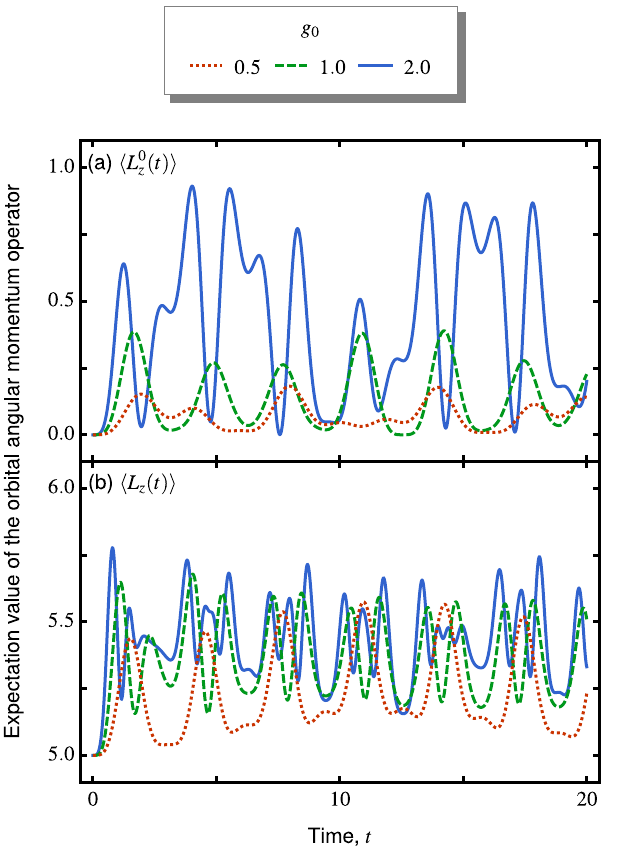}
  \caption{
    Numerical expectation value of the orbital angular momentum
    observable, using units such that $m=c=\hbar=1$ and the sinusoidal
    modulation, Eq. \eqref{eq:sin_mod}, with $\zeta=1$.
    We analyze different relativistic regimes, employing: $g_0=0.5$
    (dotted red line), $g_0=1.0$ (dashed green line) and $g_0=2.0$ (solid
    blue line). 
    In (a), we consider an initial number state with $n_r=0$, while in (b),
    we assume the initial coherent state with $|\alpha_r|^2=5$.
  }
  \label{fig:sin_l}
\end{figure}

\begin{figure}
  \centering
  \includegraphics[width=1\columnwidth]{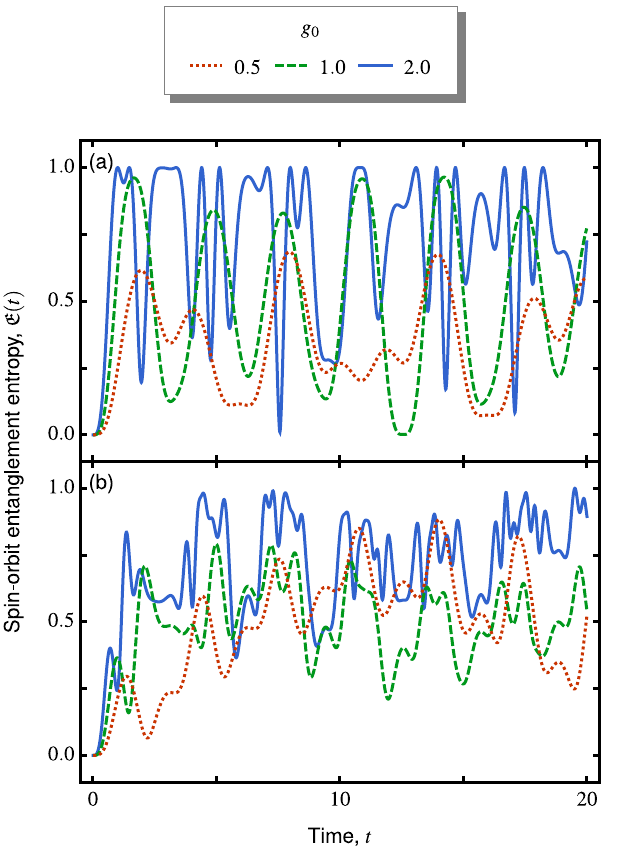}
  \caption{
    Numerical time evolution of the spin-orbit entanglement, using units
    such that $m=c=\hbar=1$ and the sinusoidal modulation,
    Eq. \eqref{eq:sin_mod}, with $\zeta=1$.
    We analyze different relativistic regimes, employing: $g_0=0.5$
    (dotted red line), $g_0=1.0$ (dashed green line) and $g_0=2.0$
    (solid blue line).
    In (a), we consider an initial number state with $n_r=0$, while in
    (b), we assume the initial coherent state with $|\alpha_r|^2=5$.
  }
  \label{fig:sin_vne}
\end{figure}

However, when analyzing numerically the DO under sinusoidal modulation
without approximation, markedly different dynamics arise.
By computing the observables and spin-orbit entanglement using the
corresponding probability amplitudes in this scenario, we obtain
aperiodic behavior, as illustrated in Figs. \ref{fig:sin_spin},
\ref{fig:sin_l}, and \ref{fig:sin_vne}.
Normally, nonlinear dynamics requires at least three coupled
differential equations to exhibit aperiodic behavior. 
However, introducing a time-dependent coupling in Eq. \eqref{eq:edos}
results in a non-autonomous system.
It can be made autonomous through an appropriate change of variables, at
the cost of increasing the system's phase space dimension
\cite{AlligoodK1996}.
This aspect of the time evolution represents an interesting approach to
the time-dependent DO problem within the emerging field of relativistic
quantum chaos \cite{Lai2018}.

This aperiodic behavior can also be achieved with other modulations for
$f(t)$ that render the system non-autonomous. 
Examples include models of atomic acceleration within the TDJC framework
\cite{Abdalla2003} or a quasiperiodic modulation of the form $f(t) =
\sin(\zeta t) \sin(\zeta' t)$ \cite{Pomeau1986}.

\section{Conclusions}
\label{sec:conc}
In this work, we have considered the time-dependent extension of the DO
from an optical perspective.
From the TDJC with time-dependent light-matter coupling, we studied the
DO with time-dependent frequency.
We presented the results considering a positive frequency in the
coupling and the resulting JC interactions for the $(2{+}1)$ dimensions
scenario.
However, we also demonstrated that the extension applies to a negative
coupling sign, as well as to the $(1{+}1)$ dimensions case for both
signs, thereby generalizing the results.
We focused our analysis on the expectation values of the angular
momentum observables and spin-orbit entanglement.
For our computations, we employ both number states and coherent states
as initial conditions.

Considering a constant coupling, we recovered the known oscillations in
the observables and entanglement for an initial number state in the
orbital angular momentum degree of freedom -- a relativistic counterpart
of the Rabi oscillations in the JC model \cite{Jaynes1963,GERRY2005},
which arise due to the phenomenon of \textit{Zitterbewegung}.
For an initial coherent state, we obtained the collapses and revivals,
as a consequence of the Poissonian distribution in the phonon number.
On the other hand, applying the exponential modulation with decaying
amplitude ($\zeta=-1$ in Eq. \eqref{eq:exp_mod}), the observables
displayed a qualitatively different behavior, reflecting the uncoupling
of the spin and orbit degrees of freedom.
The system transitions from a strongly relativistic regime to a weakly
relativistic one.
In this case, the analytical solutions were given in terms of Bessel
functions, and expressions for the asymptotic value when $\zeta<0$ were
acquired as well. 

For a trigonometric modulation in the spin–orbit coupling, we presented
the analytical solution in the Weyl oscillator limit -- mapped onto
an on-resonance JC model.
From this perspective, the behavior of the observables is dominated by
periodicity.
In contrast, for the non-approximated DO with the same time dependence, we
observed aperiodic dynamics, indicating a remarkably different time
evolution than the previous cases \cite{Prants1997}.

The simulation of the DO in cavity quantum electrodynamics was proposed
by Larson \cite{Larson2010}.  
In such setups, modern experiments in the microwave range can prepare
atomic states with different atom-field detunings
\cite{Brune1996,Sayrin2011} -- from resonance up to values on the order
of 1 MHz, where the interaction becomes minimal \cite{Brune1996}.
Additionally, the light-matter coupling may vary between 7 and 50 kHz \cite{Larson2021}.
The value mentioned above are valid for microwave configurations in
cavity quantum electrodynamics. 
Within the context of the DO, this range may be employed to explore
distinct relativistic regimes. 
Transient effects and the atomic motion along the mode shape may account
for the dynamically varying frequency of the simulated DO
\cite{Prants1992,Schlicher1989}.
In this framework, Ref. \cite{Prants1997} discusses the set of
experimental parameters that would permit observing the aperiodic
dynamics arising from the sinusoidal coupling. 
On the other hand, Pedernales \etal \cite{Pedernales2013} discussed the
prospect of relativistic simulations in circuit quantum electrodynamics.
In this platform, a time-dependent coupling appears in the form of a
tunable resonator  \cite{Maldonado-Mundo2012,Yamamoto2008,Gambetta2011,
Srinivasan2011,Yin2013,Srinivasan2014,Zeytinoglu2015}.
For trapped-ion systems, where Bermudez \etal \cite{BERMUDEZ2007}
proposed an experimental verification of the DO, a time-dependent
modulation of the parameters is also feasible, via spatially varying
optical lattices \cite{Schmiegelow2016}. 
A mapping of relevant physical quantities of the DO to possible
implementations in circuit quantum electrodynamics and trapped ions is
present in Ref. \cite{Zhang2018}. 
Therefore, we expect that the time-dependent DO studied here can be
implemented in optical systems, and we hope that our work contributes to
broadening the scope of DO verifications in analogue quantum simulators.

Furthermore, given the utility of the Dirac equation in the context of
materials science
\cite{Zhang2005,Novoselov2005,CastroNeto2009,Sadurni2011},
one might think of an inverse problem, where we control the dynamics of
the system with a given modulation in the system's parameters -- studied
in the TDJC in Refs. \cite{Yang2006,Tsutsui2025}.

\section*{Appendix: Anti-Jaynes-Cummings interaction}

We consider the DO with a negative sign in the frequency, considering
$(2{+}1)$ dimensions without loss of generality.
The corresponding Hamiltonian reads
\begin{equation} \label{eq:ajc_do_2}
    H^\prime_2 =c\sigma_{x}\left(p_x - i m\sigma_{z}\omega x \right)
    +c\sigma_{y}\left(p_{y}-i m\sigma_{z}\omega y\right)+\sigma_{z}mc^{2}.
\end{equation}

Employing the chiral operators, Eq. \eqref{eq:chiral_operators}, we
obtain the Hamiltonian in the form 
\begin{equation}
  H^\prime_2=
  \begin{pmatrix}
    mc^{2} & imc^22\sqrt{\xi}a_{l}^\dagger\\
    -imc^22\sqrt{\xi}a_{l} & -mc^{2}
\end{pmatrix},
\end{equation}
which, in the chiral quantum basis, Eq. \eqref{eq:chiral_basis}, has its corresponding energy eigenvalues given by
\begin{equation} \label{eq:ajc_energy_2+1}
    E_{2}^{\pm \prime}=\pm mc^{2}\sqrt{4\xi n_{l}+1}.
\end{equation}
That is, an energy somewhat similar to Eq. \eqref{eq:energy_2+1}, with a
subtraction of one in the orbital angular momentum quanta and the change
in chirality. 
The same can be said when we extend $\xi \to \xi(t)$, comparing with Eq. \eqref{eq:energy_2+1_time_dep}.
Employing the ladder operators, Eq. \eqref{eq:ladder}, the Hamiltonian
becomes 
\begin{equation}
  \label{eq:ajc_do_2_map}
  H_2^\prime=
  \hbar (g_2 a_{l}^{\dagger}\sigma_{+}
  + g_2^* a_{l}\sigma_{-})+mc^{2}\sigma_{z},
\end{equation}
which characterizes an AJC interaction
The conservation of angular momenta, Eq. \eqref{eq:total}, remains
valid. 
Removing the free term in Eq. \eqref{eq:ajc_do_2_map} and making
$g_2\to i g(t)$, we  have 
\begin{equation}
  \label{eq:ajc_do_2_map_time_dep}
  V^\prime(t)=  i g(t) \hbar( - a_{l}\sigma_{-}e^{-2i\theta t} +  a_{l}^\dagger\sigma_{+}e^{2i\theta t}).
\end{equation}

The subspace in this scenario results in a state in the form of
\begin{equation}
  \label{eq:ajc_state}
  \ket{\Psi^\prime(t)} = \sum_{n_l=1}^{\infty}
  \left[
    a_{n_l}(t)\ket{\uparrow,n_l} +
    b_{n_l-1}(t)\ket{\downarrow,n_l-1}
  \right].
\end{equation}
With the AJC state and the interaction picture Hamiltonian, and applying
the Schrödinger equation, we arrive at the following result: 
\begin{align}
  \dot{a}_{n_l}(t) = {}
  &
    g(t)b_{n_l-1}(t)e^{i2\theta t}\sqrt{n_l}, \\
  \dot{b}_{n_l-1}(t) = {}
  &
    -g(t)a_{n_l}(t)e^{-i2\theta t}\sqrt{n_l}.
\end{align}
The system of differential equations above can be solved in the same
ways as the ones presented previously, Eq. \eqref{eq:edos}.

\section*{Acknowledgments}
The authors thank Dr. Enrique Gabrick for helpful discussions.
This work was partially supported by Coordenação de Aperfeiçoamento de
Pessoal de Nível Superior (CAPES, Finance Code 001).
It was also supported by Conselho Nacional de Desenvolvimento
Científico Tecnológico and Instituto Nacional de Ciência e Tecnologia de
Informação Quântica (INCT-IQ).
FMA acknowledges CNPq Grant No. 313124/2023-0,
and FMA and ASMC also acknowledge
Fundação Araucária Project No. 305.

%\bibliographystyle{apsrev4-2}
%\bibliography{dirac.bib}

\end{document}